\newcommand{\be}{\begin{equation}}
\newcommand{\ee}{\end{equation}} 
\newcommand{\lb}{\label}

\newcommand{\bc}{{\bf c}}
\newcommand{\bx}{{\bf x}}
\newcommand{\bz}{{\bf z}}
\newcommand{\bh}{{\bf h}}
\newcommand{\bL}{{\bf L}}
\newcommand{\bZ}{{\bf Z}}

\documentstyle[twocolumn,prl,aps]{revtex}

\begin{document}

\relax 

\draft \preprint{} 

\title{Rayleigh-Ritz Calculation of Effective Potential 
Far From Equilibrium}

\author{Francis J. Alexander$\,^a$  and Gregory L. Eyink $\,^b$}
\address{$a$ Center for Computational Science\\
         3 Cummington Street\\
         Boston University\\
         Boston, Massachusetts 02215\\
         $\,\,\,\,\,\,$\\     
         $b$ Department of Mathematics\\
         University of Arizona\\
         Tucson, Arizona, 85721}

\date{\today}
\maketitle
\begin{abstract}
We demonstrate the utility of a Rayleigh-Ritz scheme recently 
proposed to compute the nonequilibrium 
effective potential nonperturbatively in a strong noise regime 
far from equilibrium. 
A simple 
Kramers model of an ionic conductor is used to illustrate the efficiency
of 
the method. 
\end{abstract}

\narrowtext

We have recently proposed to use in nonequilibrium statistical mechanics
a novel variational principle associated to the 
so-called {\em effective action functional} \cite{Ey96I}. This quantity 
is well-known  in quantum field theory, where the 
concept has it roots in the early work of Heisenberg \& Euler \cite{HE}
and 
Schwinger \cite{Sch} in QED. In nonequilibrium 
statistical mechanics, the first such action principle seems to have
been 
Onsager's 1931 ``principle of least dissipation'' 
\cite{O}, which applies to systems subject to thermal or molecular
noise, 
governed by a fluctuation-dissipation relation. 
A formulation of the least-dissipation principle by an action functional
on 
histories was later developed by Onsager and 
Machlup \cite{OM}. In a weak-noise limit the field-theoretic effective
action 
and the Onsager-Machlup action coincide, 
as discussed some time ago by Graham \cite{Gra}. (See also \cite{CCT}). 
However, the weak-noise limit is of rather 
restricted applicability. In particular, it is useless to deal with
problems 
in which there is no small parameter and 
in which strong fluctuations dominate the phenomena on a wide range of 
length-scales. These include, for example, high 
Reynolds-number fluid and plasma turbulence, spinodal decomposition, and
surface growth by random deposition. In strong-noise
systems, efficient calculational tools remain to be discovered. 
One powerful nonperturbative scheme which was developed 
in quantum field theory is a Rayleigh-Ritz method based upon a
constrained 
variation of the quantum energy-expectation 
functional \cite{Sym,CJT,JK}. This method has been recently extended by
us 
to nonequilibrium statistical dynamics with 
non-Hermitian evolution operator \cite{Ey96I,Ey96II}.

It is our purpose here to demonstrate the computational utility of that 
method in a simple concrete model, the {\em Kramers
model} of an ionic conductor~\cite{Kram}. The model consists of a unit
mass, 
charged particle in a 1-dimensional cosine 
potential $V(x)=V_0\cos x$, damped by linear friction with coefficient 
$\gamma$, and driven both by an electric force $F=qE$ 
and by thermal noise of temperature $T=\gamma\Theta/k_B$. The dynamics
of 
individual realizations is given by the nonlinear 
Langevin equation 
\be \ddot{x}+\gamma\dot{x}+V_0\sin x=F+\Gamma(t), \lb{Keq} \ee
with zero-mean Gaussian noise:
\be \langle\Gamma(t)\Gamma(t')\rangle=2\gamma\Theta\delta(t-t'). 
\lb{cov} \ee
Here $x$ runs over $[0,2\pi]$ with periodic b.c. Equivalently, the 
evolution of the single-time distribution
$P_t$ of position $x$ and velocity $v=\dot{x}$ is given by the 
Fokker-Planck equation 
\begin{eqnarray}
{{\partial}\over{\partial t}} 
P_t & = &-{{\partial}\over{\partial x}}\left(v P_t\right)
      +{{\partial}\over{\partial v}}[(\gamma v+V_0^{\,}\cos x \cr
  \, & & \,\,\,\,\,\,\,\,\,\,\,\,\,\,\,\,\,\,\,\,\,\,\,\,\,\,\,\,\,\,\,
-F+\gamma\Theta{{\partial}\over{\partial v}})P_t]\equiv \hat{L}P_t. 
\lb{FPeq}
\end{eqnarray}
The important point for our discussion here is that all parameter values
will be 
chosen order one, i.e.
$V_0=\gamma=F=\Theta=1.$ In this case, there is neither a small nor a
large 
parameter on which to base a 
perturbation expansion or asymptotic evaluation. In particular, $F=1$
implies 
that the system is far from thermal 
equilibrium, well beyond the regime of linear response of steady-state 
current $j(F)=\langle v\rangle_F$ to the applied 
field  $F$. 

In the statistical steady state, rather than the full effective-action, 
it is more useful to consider its time-extensive 
limit, the  so-called {\em effective potential}. For any random variable
$\bz(t)$ in an ergodic system, this 
quantity measures the ``cost'' for fluctuations to occur in  the
empirical 
time-averages: 
\be \overline{\bz}_T\equiv {{1}\over{T}}\int_0^Tdt\,\,\bz(t). 
\lb{empavrg} \ee
That is, it measures the probability for deviations to occur in a given 
realization at large $T$ between 
$\overline{\bz}_T$ and the ensemble-mean $\overline{\bz}.$
Quantitatively, the relation between this probability and 
the effective potential function $V$ is that, for any possible
single-time value $\bz$ of the variable 
$\bz(t)$, the probability
for the average over a time-interval of duration $T$ to take on the 
value $\bz$ has magnitude
\be {\rm Prob}\left(\{\overline{\bz}_T\approx \bz\}\right)\sim
\exp\left(-T\cdot V [ \bz ] \right) \lb{fluc} \ee
in the limit of large $T$. The potential function $V[\bz]$ is
nonnegative 
and convex, with minimum value =0 occuring only 
at the ensemble-mean, $V[\overline{\bz}]=0.$  The fluctuation formula 
Eq.(\ref{fluc}) is a refinement of the standard 
ergodic hypothesis. It states not only that time-averages
$\overline{\bz}_T$ 
will converge to the ensemble-average 
$\overline{\bz}$ almost surely in the limit of large $T$, but also it 
gives a precise quantitative estimate on the 
probability of the deviations. This refinement holds whenever some 
condition of finite exponential moments
is satisfied: see \cite{Ey96I,Ey96II}. The effective potential is 
analogous to the thermodynamic entropy of an 
equilibrium system (more precisely, to its negative) and Eq.(\ref{fluc})
for the probability of statistical deviations 
is analogous to the Einstein-Boltzmann formula for the fluctuations away
from equilibrium. This probabilistic 
interpretation of the effective potential was pointed out in
field-theory 
by Jona-Lasinio \cite{GJL}. Although any 
variable might be considered, we shall be interested here in the 
electric current $j$ and its effective potential $V[j]$.

The method we use to calculate $V$ is the Rayleigh-Ritz variational
scheme 
proposed in \cite{Ey96I,Ey96II}. As discussed 
in more detail there, the effective potential may be characterized very 
generally by means of a constrained variation 
of the ``energy functional'' 
\be {\cal H}[\Psi^R, \Psi^L]\equiv -\langle\Psi^L,\hat{L}\Psi^R\rangle, 
\lb{enerfun} \ee
defined in terms of the Liouville operator $\hat{L}$ of the
nonequilibrium statistical 
dynamics. $V[\bz]$ was
shown to be the extremal value of ${\cal H}[\Psi^R, \Psi^L]$ under
arbitrary 
variation of ``trial states'' $\Psi^{R,L}$ 
subject to the constraints of constant overlap 
\be \langle\Psi^L,\Psi^R\rangle=1 \lb{ovlap} \ee
and constant means
\be \langle\Psi^L,\hat{\bZ}\Psi^R\rangle=\bz. \lb{constmn} \ee
The operator $\hat{\bZ}$ consists of multiplication by variable $\bz.$ 
By means of Lagrange multipliers $\lambda,\bh$
associated to the constraints, this variational characterization of the
effective 
potential can be expressed in spectral 
terms as
\be V[\bz]= \bz\cdot\bh-\lambda[\bh], \lb{Legtrans} \ee 
where $\lambda[\bh]$ is the principal eigenvalue of the ``perturbed'' 
Liouville operator 
$\hat{L}_\bh=\hat{L}+\bh\cdot\hat{\bZ}.$ Observe that this spectral
characterization 
is exactly analogous to
the characterization of the free-energy in equilibrium statistical
mechanics 
as the principal eigenvalue of the
``transfer-matrix'' in lattice systems\cite{TrMtrx}. In that case, the
entropy is then obtained by a 
Legendre transform of 
the free-energy, entirely analogous to the relation of $\lambda[\bh]$
and 
$V[\bz]$ in Eq.(\ref{Legtrans}) by
a Legendre transform. As the starting point for an approximate
Rayleigh-Ritz calculation of $V$, any probability 
distribution function (PDF) ansatz for the 
stationary distribution of the nonequilibrium dynamics may be employed. 
The ``right trial state'' $\Psi^R(\bx)$ just consists
of a guess for the stationary PDF, $\rho(\bx;\bc)$, depending upon some 
arbitrary parameters $\bc.$ The ``left trial state''
$\Psi^L(\bx)$ is a (linear superposition of a) set of selected
moment-functions $\psi_n(\bx).$ 
The variation within such a
framework leads in general to a ``nonlinear eigenvalue problem'' to
determine 
the approximate value $\lambda_{\#}[\bh]$ of 
the leading eigenvalue within the ansatz. See \cite{Ey96II}.  

The simplest systematic approximation procedure in the Kramers model is 
to use the Gaussian trial weight
\be w(x,v)={{1}\over{\sqrt{2\pi\Theta}}}\exp\left[-{{v^2}
\over{2\Theta}}\right], \lb{trwt} \ee
along with the corresponding truncated orthogonal polynomial expansions
\be \Psi^R(x,v) = w(x,v)\cdot\sum_{n=0}^N\sum_{p=-P}^P 
c_{n,p}^R\cdot {{He_n(v)}\over{\sqrt{n!}}}\cdot{{e^{ipx}}
\over{\sqrt{2\pi}}} \lb{Rexp} \ee
and
\be \Psi^L(x,v) = \sum_{n=0}^N \sum_{p=-P}^P 
c_{n,p}^L\cdot {{He_n(v)}\over{\sqrt{n!}}}\cdot{{e^{ipx}}
\over{\sqrt{2\pi}}} \lb{Lexp} \ee 
Here $He_n(v)$ are a set of Hermite polynomials, with conventions as in
\cite{AS}. 
The same expansion was 
used by Risken \& Vollmer \cite{RV} to calculate the full stationary
measure. 
Orthogonal polynomial expansions have the 
simplifying feature that they lead to a linear problem for the
approximate
 eigenvalue $\lambda_{\#}[h]$ and 
eigenvectors $\bc^R_{\#},\bc^L_{\#}$ (with $\#$ shorthand for $N,P$), as
\be \sum_{n',p'}(\bL_{\#,h})_{np,n'p'}c_{\#,n'p'}^R= 
\lambda_{\#}[h]c_{\#,np}^R, \lb{Reigeq} \ee
and 
\be \sum_{n',p'}(\bL_{\#,h}^\dagger)_{np,n'p'}c_{\#,n'p'}^L=
\lambda_{\#}^*[h]c_{\#,np}^L, 
\lb{Leigeq} \ee
with 
\begin{eqnarray}
\, & & (\bL_{\#,h})_{np,n'p'}= \sqrt{(n+1)\Theta}(-ip+h)\delta_{pp'}
\delta_{n,n'-1} \cr
\, & & \,\,\,\,\,\,\,\,\,\,\,\,\,\,\,\,\,\,\,\,\,\,\,\,\,\,\,\,\,\,\,\,\,
       \,\,\,\,\,\,\,\,\,\,\,\,\,\,\,\,\,\,\,\,\,\,\,\,\,\,\,\,\,\,\,\,\,
       \,\,\,\,\,\,\,\,\,\,\,\,\,\,\,\,\,\,\,\,\,\,\,
        -n\gamma\delta_{pp'}\delta_{nn'} \cr
\, & & +\sqrt{n\Theta}\left[\left(-ip+h+{{F}\over{\Theta}}\right)
\delta_{pp'} \right.\cr
\, & & \,\,\,\,\,\,\,\,\,\,\,\,\,\,\,\,\,\,\,\,\,\,\,\,\,\,\,\,\,\,\,\,\,
       \left.+i{{V_0}\over{2\Theta}}\left(\delta_{p,p'+1}-
\delta^{p,p'-1}\right)\right]\delta_{n,n'+1}, \lb{Lhmat} 
\end{eqnarray}
for $n=0,...,N, p=-P,...,P$ and
$(\bL_{\#,h}^\dagger)_{np,n'p'}=(\bL_{\#,h})_{n'p',np}^*$, 
the Hermitian conjugate.
$\bL_{\#,h}$ is the finite matrix approximation to the full
``perturbed'' operator 
$\hat{L}_h=\hat{L}+h\hat{V}.$ 
The approximate value of the effective potential is then obtained from 
\be V_{\#}[j]= j_{\#}[h]\cdot h-\lambda_{\#}[h], \lb{epot} \ee
in which  $\lambda_{\#}[h]$ is the eigenvalue branch passing through $0$
at $h=0$ 
and $j_{\#}[h]\equiv \langle\Psi_{\#}^L(h),
\hat{V}\Psi_{\#}^R(h)\rangle$ is the associated ``perturbed''
current. The latter 
may be determined from the 
Hellmann-Feynman theorem\cite{Merz}, as
$j_{\#}[h]=\lambda_{\#}^\prime[h].$ It can 
also be calculated directly from the approximate 
right and left eigenvectors, as 
\begin{eqnarray}
j_{\#}[h] & = & \sum_{n,p}\left[\sqrt{n}(c_{\#,n-1,p}^L)^*c_{\#,n,p}^R 
\right. \cr
\, & & \,\,\,\,\,\,\,\,\,\,\,\,\,\,\,\,\,\,\,\,\,\,\,\,\,\,\,\,\,\,\,\,\,
                     \left. +\sqrt{n+1}(c_{\#,n+1,p}^L)^*c_{\#,n,p}^R
\right] .    
\lb{jeq} 
\end{eqnarray}
This avoids the evaluation of derivatives and is more computationally
efficient in 
such a small system, where the 
determination of the eigenvalues and eigenvectors is easy to
accomplish. These 
calculations have been 
carried out by us numerically, and the results for the approximate
effective 
potential $V_{\#}[j]$ are graphed 
in {\em Figure 1} for various choices of $N$ and $P$. As can be seen
from that 
figure, convergence is already obtained 
in the range $j=0.2-1.4$ to 1\% accuracy for $N=4,P=5$. 

The effective potential can also be obtained by direct numerical
solution (DNS) 
of the Langevin dynamics. The most obvious 
procedure would be to gather long time series of $v(t)$ in the
steady-state, to 
assemble histograms of probabilities of 
empirical time averages $\overline{v}_T={{1}\over{T}}\int_0^T
dt\,\,v(t)$ over 
time-intervals of duration $T$, and then to 
calculate $V[j]$ via the inverse to Eq.(\ref{fluc}). However, we have
found that 
the approximate potential from this method 
converges very poorly. Instead, the most efficient method is to
determine 
approximate values of $\lambda_\#[h]$ 
and $j_\#[h]$ from
\be \lambda_\#[h]={{1}\over{T}}\ln\langle\exp\left[h\int_0^T dt\,\,v(t)
\right]\rangle \lb{lamdirdef} \ee
and 
\be j_\#[h]= {{\langle\overline{v}_T\exp\left[h\int_0^T
    dt\,\,v(t)\right]
\rangle}
               \over{\langle\exp\left[h\int_0^T
                 dt\,\,v(t)\right]\rangle}}, \lb{jdirdef} \ee
where $\langle\cdot\rangle$ denotes average over some number $R$ of
realizations 
and $\#$ now stands for $R$, $T$
and the numerical time step $\Delta t$. Substituting these into
Eq.(\ref{epot}), 
a value $V_\#[j]$ is obtained 
which for large $R,T$ and small $\Delta t$ converges to the true
potential. 
We have also performed this direct 
calculation of $V[j]$ via a 2nd-order stochastic Runge-Kutta integration
of the 
Langevin Eq.(\ref{Keq}) for $R=32000$,
$T=1000$ and $\Delta t=0.01$. 
Increasing $T$ or decreasing $\Delta t$ did not change the results, for 
which the largest
errors are statistical, based upon the finite number of realizations $R$. 

In {\em Figure 2} we compare also the potentials $V[j]$ 
themselves for the above DNS and the $N=10,P=8$ Rayleigh-Ritz, over 
the range $j=0.718-0.733$. 
Over this range the error in the Rayleigh-Ritz value is 
at most one part in $10^4$. The 
relatively large error bars on $V_{{\rm DNS}}$ are due to the
cancellation 
of linear terms in $h$ between $\lambda_\#[h]$ and 
$j_\#[h]\cdot h$ in Eq.(\ref{epot}). Because of this cancellation, 
getting an accuracy even to one part in $10^2$ for $V_\#$ 
required an accuracy to one part in $10^4$ for $\lambda_\#[h]$ 
and $j_\#[h]$. 
To achieve an accuracy of $V_{DNS}$ comparable to Rayleigh-Ritz would
require a reduction in statistical error in DNS values of $\lambda_{\#}$
and $j_\#$ from 1 part in $10^4$ to 1 part in $10^8$.  
With this error estimated as 
$O(R^{-1/2})$, it  would require a computation time longer by a factor
of $10^8$.
The present calculation required
 $\sim 10^{12}$ floating-point operations 
(flops) and, on the machine we employed, took about 10 hrs, the bulk of
which was devoted to generating random numbers.  Thus, accuracy to one 
part in $10^4$ for $V_\#$ from DNS would 
require about $10^5$ years of computation! By contrast, the Rayleigh-Ritz 
computation with $N=10,P=8$ accurate to $0.01\%$ 
required $\sim  10^6$ flops, performed in about $0.1$ seconds. 
In general, the Rayleigh-Ritz evaluation 
of $V_\#[j]$ in the Kramers model requires for each value of 
$h$ a calculation of the leading eigenvalue $\lambda_\#[h]$ 
and its associated eigenvector for the tridiagonal matrix 
Eq.(\ref{Lhmat}), which is an $O(N\cdot P)$ operation, and the 
calculation of current $j_\#[h]$ via the summation formula 
Eq.(\ref{jeq}), which is also $O(N\cdot P)$. The number of flops 
required is $\sim 1000 N\cdot P$. The superiority of the Rayleigh-Ritz 
method compared with DNS is clear. Even more favorable
is the fact that the Rayleigh-Ritz method gives a result 
as in {\em Figure 1} over a range 100 times larger, to $1\%$ 
accuracy, with only $N=4,P=5$.  In this range errors in DNS 
due to finite $T$ and $R$ are very large because the 
$h$-weighted ensembles in Eqs.(\ref{lamdirdef}),(\ref{jdirdef}) 
have a greater contribution from rare events as $h$ 
increases. Outside of a narrow range of current $j$
near the mean value the determination of effective potential by DNS 
is practically impossible.   

In this work we have focused upon the Rayleigh-Ritz determination of the
effective potential via a convergent 
scheme, the expansion in orthogonal polynomials. However, as discussed
in 
detail in \cite{Ey96II} any physically-inspired, 
nonlinear ansatz for the PDF or any ``surrogate'' random variables
chosen 
to model the system variables may be used as 
well. This is important for systems with many degrees-of-freedom, such
as 
high Reynolds number turbulence or large-scale 
dynamics of multiphase fluids, in which convergent schemes are presently
not remotely feasible. In work in progress 
\cite{AlEy}, we apply our variational methods to practical modeling of 
such systems, in particular hydrodynamic 
turbulence. Analogous time-dependent Rayleigh-Ritz methods are available
to calculate the full effective action 
functional on histories: see \cite{Ey96II}. There is some general
similarity of 
our methods to the Hartee-Fock 
variational approximation applied by Crisanti \& Marconi \cite{CM} to
the calculation 
of effective action in phase 
segregation dynamics. Both are intrinsically nonperturbative and capable
of systematic 
improvement. In addition, 
the Rayleigh-Ritz scheme illustrated here has a very great flexibility
to incorporate 
intuitive guesses into an 
analytical calculation: any PDF ansatz for the system variables
whatsoever may be 
used as the basis for a calculation 
by our variational method. 

\noindent {\bf Acknowledgements}
We would like to thank Y. Oono for useful discussions on this subject.
Numerical simulations were carried out at the Center for Computational
Science at Boston University and the Department of Mathematics at the 
University of Arizona.

\newpage
\begin{center}
FIGURE CAPTIONS
\end{center}

Figure 1.)  Approximate effective potential $V[j]$
for N=4, P=5 ($ \Box$),
N=6, P=6 ($ + $) and
N=10, P=8 ($ \diamond$).

Figure 2.) 
Potentials $ V[j] $ for DNS (with errorbars) and the $N=10,P=8$
Rayleigh-Ritz solid line).

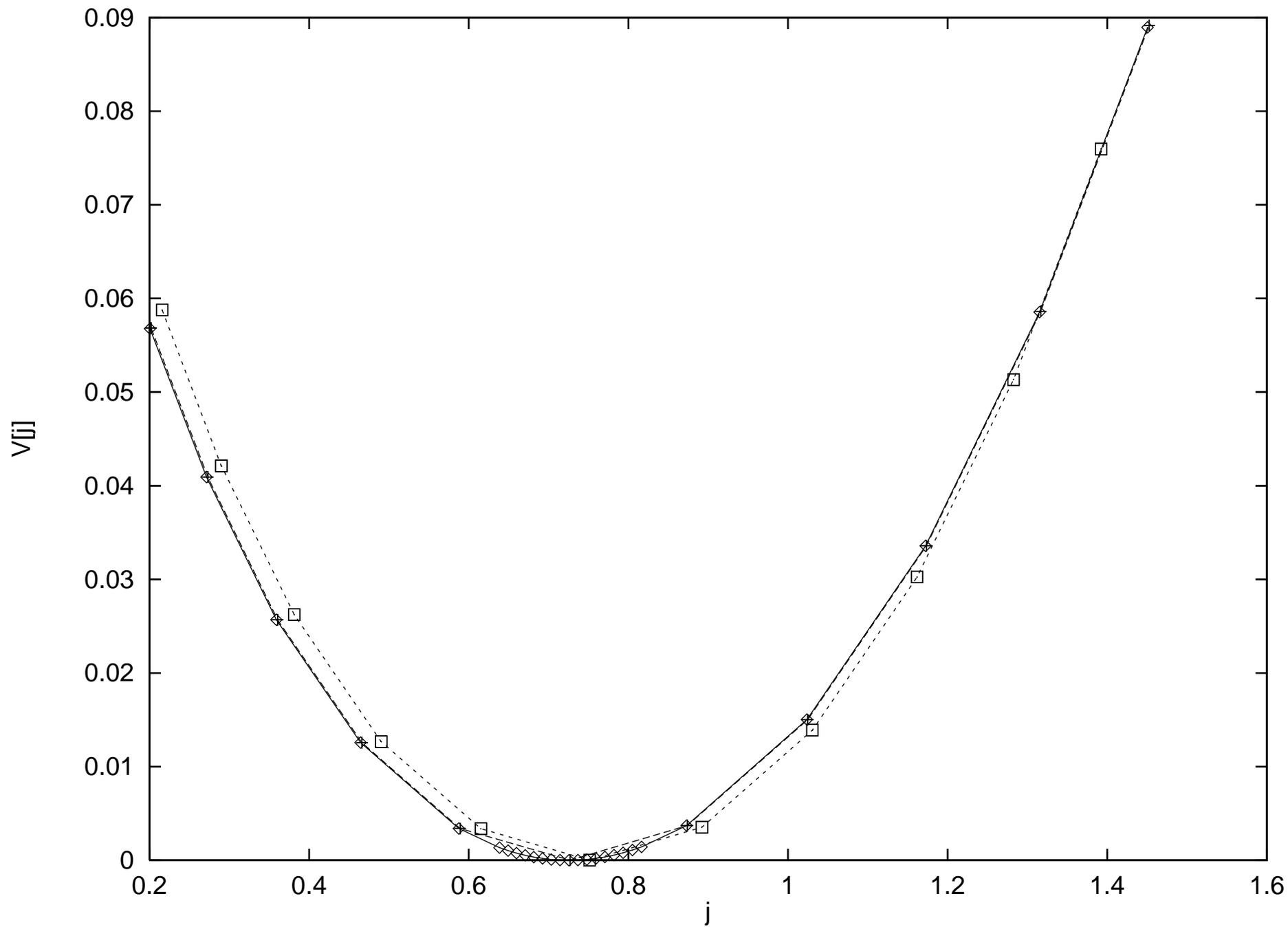
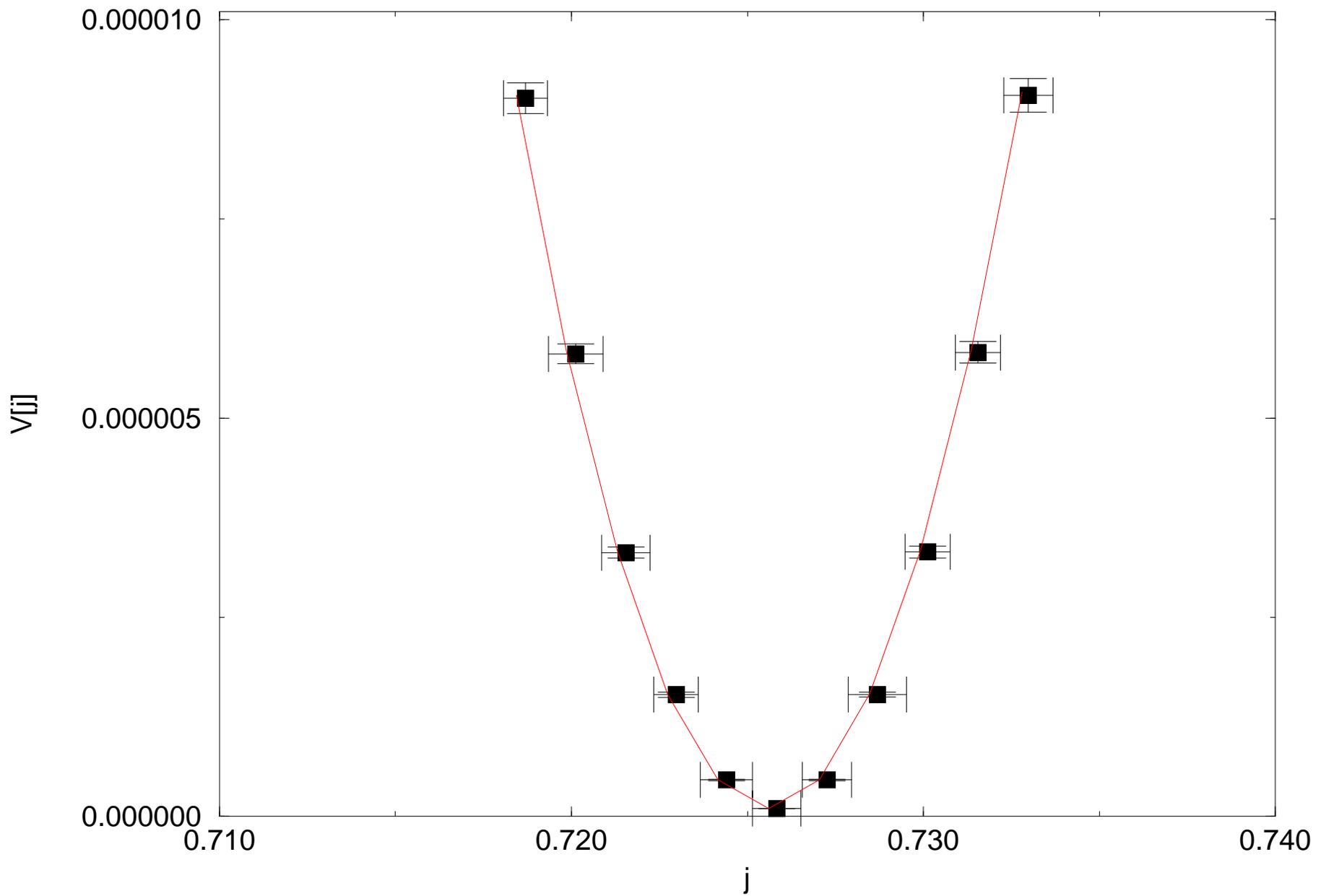
\end{document}